# Mechanical behavior, bonding nature and defect processes of $Mo_2ScAlC_2$: a new ordered MAX phase

M. A. Hadi[1,a)], S. H. Naqib[1], S.-R. G. Christopoulos,[2] A. Chroneos,[2,3,b)], and A. K. M. A. Islam[1,4]

[1]Department of Physics, University of Rajshahi, Rajshahi-6205, Bangladesh
[2]Faculty of Engineering, Environment and Computing, Coventry University, Priory Street, Coventry CV1 5FB, UK
[3]Department of Materials, Imperial College, London SW7 2AZ, UK
[4]International Islamic University Chittagong, 154/A College Road, Chittagong 4203, Bangladesh

**Abstract**

In the present study we employed density functional theory calculations to investigate the mechanical behavior, bonding nature and defect processes of the new ordered MAX phase $Mo_2ScAlC_2$. The mechanical stability of the compound is verified with its single crystal elastic constants. The new phase $Mo_2ScAlC_2$ is anticipated to be prone to shear along the crystallographic *b* and *c* axes, when a rational force is applied to the crystallographic *a* axis. The compressibility along the ⟨001⟩ direction under uniaxial stress is expected to be easier in $Mo_2ScAlC_2$. Additionally, the volume deformation should be easier in $Mo_2ScAlC_2$ than the isostructural $Mo_2TiAlC_2$. $Mo_2ScAlC_2$ is predicted to behave in a brittle manner. Due to its higher Debye temperature, $Mo_2ScAlC_2$ is expected to be thermally more conductive than $Mo_2TiAlC_2$. The cross-slip pining procedure should be significantly easier in $Mo_2ScAlC_2$ as compared to $Mo_2TiAlC_2$. The new ordered MAX phase $Mo_2ScAlC_2$ has a mixed character of strong covalent and metallic bonding with limited ionic nature. Both Mo–C and Mo–Al bonds are expected to be more covalent in $Mo_2ScAlC_2$ than those of $Mo_2TiAlC_2$. The level of covalency of Sc–C bond is somewhat low compared to a similar bond Ti–C in $Mo_2ScAlC_2$. Due to its reduced hardness $Mo_2ScAlC_2$, it should be softer and more easily machinable compared to $Mo_2TiAlC_2$. Fermi surface topology of the new compound is formed mainly due to the low-dispersive Mo 4d-like bands. The intrinsic defect processes reveal that the level of radiation tolerance in $Mo_2ScAlC_2$ is not as high as in other MAX phases such as $Ti_3AlC_2$.

# 1. Introduction

A class of layered laminated ternary compounds, known as MAX phases with chemical formula $M_{n+1}AX_n$ (M = early transition metal from group 3 – 6, A = element from columns 12 – 16 in the periodic table and X = C and/or N) has gained the interest of the community due to their use and potential for technological applications. For $n$ = 1, 2, 3, the sub-families of these compounds are classified as 211, 312, 413 MAX phases, respectively. In 1960s, Nowotny *et al.* [1] first discovered some members of 211 MAX phases and named them H-phases [1]. MAX phases were revived in 1995 when Barsoum and El-Raghy [2] discovered $Ti_3AlC_2$ and demonstrated its unique combination of physical and chemical properties inherent to the compounds in the MAX family. These compounds adopt the hexagonal $P6_3/mmc$ structure in which M and X atoms form octahedral edge-sharing building blocks interleaved by A-atomic layers. This structure is highly anisotropic and catalytic for the combination of metallic and ceramic characteristics in MAX phases [3]. The common metallic properties characterized by MAX phases are thermal and electrical conductivities, resistance to thermal shock, plasticity at high temperature, damage tolerance and machinability [2,4–7]. The ceramic-like properties displayed by MAX phases are elastic rigidity, lightweight, creepiness, fatigue, resistance to oxidation and corrosion, maintaining the strength to high temperatures [8–13]. Due to these technologically important properties, MAX phases are already employed in the fields of defense, aerospace, medical, automobile, portable electronics and nuclear reactor [14,15].

About 70 MAX phases have been synthesized in bulk form amongst which only $Mo_2GaC$, $Mo_2TiAlC_2$ and $Mo_2Ti_2AlC_3$ contain Mo as the M element. Very recently, $Mo_2ScAlC_2$ was synthesized by heating the mixture of elemental powders of Mo, Sc, Al and graphite at 1700°C [16]. The high resolution transmission electron microscopy (HRTEM) study ensures that $Mo_2ScAlC_2$ is a chemically ordered structure with one Sc layer sandwiched between two Mo-C layers [16]. However, there is only one Sc-based MAX phase $Sc_2InC$ synthesized until now [17,18]. Therefore, the newly synthesized $Mo_2ScAlC_2$ is an exceptional member in MAX family that combines two uncommon transition metals Mo and Sc as an M element. In the present study, we aim to explore the mechanical and bonding properties of $Mo_2ScAlC_2$ for the first time including intrinsic defect processes. Additionally, we compare it with the isostructural $Mo_2TiAlC_2$. The paper is arranged in three sections. In Section 2, a concise description of the computational methodology is presented. The results obtained for the structural, elastic, electronic properties, and intrinsic defect processes of $Mo_2ScAlC_2$ are analyzed in Section 3. Finally, Section 4 summarizes the main conclusions of the present investigation.

## 2. Computational methods

The plane wave pseudopotential method within the density functional theory (DFT) as implemented in the CASTEP code [19] is used. The generalized gradient approximation based on Perdew-Burke-Ernzerhof functional (GGA–PBE) [20] is employed to describe the exchange-correlation energy. The Vanderbilt type ultrasoft pseudopotential is chosen to treat the electron-ion interactions [21]. A plane-wave energy cut-off is set as 550 eV for expanding the plane wave functions. The first Brillouin zone is sampled with a 17×17×2 $k$-point mesh according to Monkhorst-Pack (MP) scheme [22]. Broyden-Fletcher-Goldfarb-Shanno (BFGS) algorithm [23] is applied to relax the structure fully with respect to atomic positions and lattice parameters. The tolerance for energy, maximum force, maximum stress, and maximum atomic displacement are chosen to be within $5\times10^{-6}$ eV/atom, 0.01 eV/Å, 0.02 GPa and $5\times10^{-4}$ Å, respectively. To obtain smooth Fermi surface, the electronic structures are calculated with an energy cut-off of 600 eV and 25×25×3 $k$-point mesh.

The elastic constants can be calculated by using the finite-strain theory formulated within CASTEP module [24]. In this method, a set of specified uniform deformations (strains) of finite value is applied and after that the consequential stress is evaluated with respect to optimizing the internal degrees of freedom. This method has been used to adequately calculate the elastic properties of many materials together with metallic systems [25–34]. The stress tensor $\sigma_{ij}$ under a set of applied strain $\delta_j$ gives the elastic constants via the equation,

$$\sigma_{ij} = \sum_{ij} C_{ij}\delta_j \tag{1}$$

The polycrystalline bulk elastic properties are calculated within the scheme of the Voight-Reuss-Hill (VRH) approximation for polycrystalline aggregates [35–37] by using the calculated elastic constants. The Young's modulus and Poisson's ratio are also determined from the relations $Y = (9GB)/(3B + G)$ and $v = (3B – 2G)/(6B+ 2G)$, respectively.

The calculations for the intrinsic defect processes involved a 108-atomic site supercell (under constant pressure conditions) and a 3 x 3 x 1 MP $k$-point grid. For the interstitial sites we performed a comprehensive investigation.

## 3. Results and discussions
### 3.1. Structural properties

$Mo_2ScAlC_2$ crystallizes in a hexagonal structure with the space group of $P6_3/mmc$ like all other MAX phases and is isostructural with $Mo_2TiAlC_2$. The crystal structure of $Mo_2ScAlC_2$ is represented in Fig. 1. The unit cell of $Mo_2ScAlC_2$ contains 12 atoms and two formula units (Z = 2). Table 1 lists the calculated lattice constants $a$ and $c$ as well as equilibrium unit cell volume together with atomic positions. The theoretical results are in good agreement with the values obtained in the experiment with deviations for lattice constants $a$ and $c$ and unit cell volume $V$ as 0.63, 1.54 and 2.82%, respectively. It can be observed from Table 1 that the exchange of Ti with Sc in $Mo_2TiAlC_2$ causes an increase in lattice constants and unit cell volume.

**Table 1**: Structural properties obtained from DFT based first-principles calculations along with corresponding experimental data.

| Structural Properties | | Experimental data | | Theoretical data | |
|---|---|---|---|---|---|
| | | $Mo_2ScAlC_2$ [16] | $Mo_2TiAlC_2$ [38] | $Mo_2ScAlC_2$ [This] | $Mo_2TiAlC_2$ [31] |
| $a$ (Å) | | 3.0334 | 2.997 | 3.0524 | 2.998 |
| $c$ (Å) | | 18.7750 | 18.661 | 19.0638 | 18.751 |
| $V$ (Å$^3$) | | 149.6132 | 145.157 | 153.8285 | 145.955 |
| Space group | | $P6_3/mmc$ | | | |
| Atomic position (Rietveld) | | | | | |
| Mo | 4f | (1/3, 2/3, 0.13632) | (1/3, 2/3, 0.13336) | (1/3, 2/3, 0.13716) | (1/3, 2/3, 0.13316) |
| Sc/Ti | 2a | (0, 0, 0) | (0, 0, 0) | (0, 0, 0) | (0, 0, 0) |
| Al | 2b | (0, 0, 1/4) | (0, 0, 1/4) | (0, 0, 1/4) | (0, 0, 1/4) |
| C | 4f | (2/3, 1/3, 0.06825) | (2/3, 1/3, 0.06875) | (2/3, 1/3, 0.07619) | (2/3, 1/3, 0.06857) |

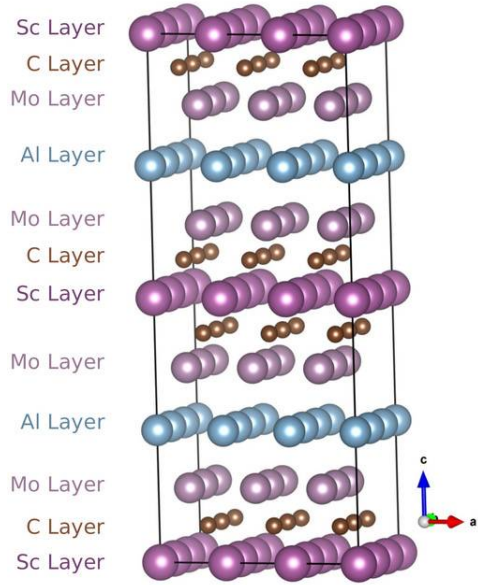

Fig. 1. Crystal structure of chemically ordered MAX phase, $Mo_2ScAlC_2$.

*3.2. Elastic properties*

The mechanical behavior of solids is dependent upon their elastic constants. In particular, elastic constants of solids are significantly related to the bonding characteristics between adjacent atomic planes and the anisotropic nature of the bonding and structural stability. Due to its hexagonal crystal structure, the new ordered MAX phase $Mo_2ScAlC_2$ has six different elastic constants namely, $C_{11}$, $C_{12}$, $C_{13}$, $C_{33}$, $C_{44}$, and $C_{66}$. Only five of them are independent in view of the fact that $C_{66} = (C_{11} - C_{12})/2$. These independent elastic tensors determine the mechanical stability of the materials. For instance, the mechanical stability of the crystals with hexagonal structure can be assessed with the independent elastic constants via the conditions [39]:

$$C_{11} > 0, \quad C_{33} > 0, \quad C_{44} > 0, \quad (C_{11} - C_{12}) > 0, \text{ and } (C_{11} + C_{12})C_{33} > 2C_{13}^2 \quad (2)$$

The calculated elastic constants of $Mo_2ScAlC_2$ are listed in Table 2. All these constants are positive and the above conditions are satisfied, illustrating the mechanical stability of the newly fabricated $Mo_2ScAlC_2$.

**Table 2**. Calculated single crystal elastic constants $C_{ij}$ (GPa), polycrystalline bulk modulus $B$ (GPa), shear modulus $G$ (GPa), Young modulus $Y$ (GPa), Pugh's ratio $G/B$, Poisson's ratio $v$ and shear anisotropy factor $A$ of $Mo_2ScAlC_2$ in comparison with $Mo_2TiAlC_2$

| Compounds | Single crystal elastic properties | | | | | | Polycrystalline elastic properties | | | | | | Ref. |
|---|---|---|---|---|---|---|---|---|---|---|---|---|---|
| | $C_{11}$ | $C_{12}$ | $C_{13}$ | $C_{33}$ | $C_{44}$ | $C_{66}$ | $B$ | $G$ | $Y$ | $B/G$ | $v$ | $A$ | $k_c/k_a$ | |
| $Mo_2ScAlC_2$ | 293 | 109 | 117 | 290 | 134 | 92 | 173 | 105 | 262 | 1.65 | 0.25 | 1.54 | 0.97 | This |
| $Mo_2TiAlC_2$ | 393 | 132 | 133 | 371 | 160 | 131 | 217 | 140 | 346 | 1.55 | 0.23 | 1.29 | 1.09 | [31] |
| | 386 | 143 | 140 | 367 | 150 | 131 | 220 | 131 | 329 | 1.68 | 0.25 | 1.27 | 1.10 | [40] |

The elastic stiffness of a material regarding the (100)⟨100⟩ uniaxial strain can be estimated by its elastic constant $C_{11}$. It is observed that the replacement of Ti with Sc decreases the stiffness of $Mo_2ScAlC_2$. It is, therefore, expected that the new phase $Mo_2ScAlC_2$ should be softer and more easily machinable than $Mo_2TiAlC_2$. The independent elastic constants $C_{12}$ and $C_{44}$ are linked to the elasticity in shape. In particular, the elastic tensor $C_{12}$ is related to the pure shear stress in the (110) plane along the ⟨100⟩ direction, whereas the elastic constant $C_{44}$ is caused by the shear stress in the (010) plane in the ⟨001⟩ direction. The shear deformations associated with these two shear stresses are expected to be easier in $Mo_2ScAlC_2$ than in $Mo_2TiAlC_2$. The elastic constants $C_{12}$ and $C_{13}$ combine a functional stress component along the crystallographic $a$ direction in the presence of a uniaxial strain along the crystallographic $b$ and $c$ axes, uniformly. Due to low values of these constants in comparison with those of $Mo_2TiAlC_2$, the new compound $Mo_2ScAlC_2$ is expected to be prone to shear along the crystallographic $b$ and $c$ axes, when a reasonable force is applied to the crystallographic $a$ axis. The independent elastic tensor $C_{33}$ is due to the uniaxial deformation along the ⟨001⟩ direction. The substitution of Ti from $Mo_2TiAlC_2$ with Sc causes a decrease of $C_{33}$ and consequently, the compressibility along the ⟨001⟩ direction under uniaxial stress is expected to be easier in $Mo_2ScAlC_2$.

The bulk elastic parameters are calculated and listed in Table 3. The bulk modulus $B$ measures the aptitude of solids to resist compression. It also reflects the nature of chemical bonding within a solid. The replacement of Ti with Sc causes a significant decrease in bulk modulus of $Mo_2ScAlC_2$. Therefore, the volume deformation is expected to be easier in $Mo_2ScAlC_2$ than in $Mo_2TiAlC_2$. Moreover, the strength of chemical bonding in $Mo_2ScAlC_2$ is low as compared to $Mo_2TiAlC_2$. All these signify that the new phase $Mo_2ScAlC_2$ is softer than $Mo_2TiAlC_2$. The shear modulus $G$ assesses the materials' ability to resist their plastic deformation i.e., shape change. It also correlates with hardness and the elastic constant $C_{44}$. It is observed that the shear modulus decreases significantly when the Ti atom is replaced with Sc from $Mo_2TiAlC_2$. Therefore, the shape change in $Mo_2TiAlC_2$ should not be as easy as in $Mo_2ScAlC_2$, which also implies that $Mo_2ScAlC_2$ is more easily machinable than $Mo_2TiAlC_2$. The Young's modulus $Y$ is a measure of the stiffness of a solid material, with a stiff material requiring more force to deform as compared to a soft material. A remarkable decrease is observed in Young's modulus when comparing $Mo_2TiAlC_2$ to $Mo_2ScAlC_2$, indicating that in view of Young's modulus the new compound $Mo_2ScAlC_2$ should also be softer than $Mo_2TiAlC_2$.

The failure mode of solids (brittle or ductile failure) can be explained depending on their bulk and shear modulus. This is of importance to determine the integrity of the structure. A sudden fracture appears in a material due its brittle failure; conversely plastic deformation occurs before fracture for a material that undergoes ductile failure. In ductile failure, the crack progresses slowly with a large amount of plastic deformation and it will typically not spread unless an increased stress is applied. Conversely, in brittle failure, cracks extend very rapidly with little or no plastic deformation. Pugh [41] used the bulk to shear modulus ratio $B/G$ as a parameter to determine whether a material is brittle or ductile. The critical value of 1.75 for $B/G$ is the borderline between ductile and brittle materials [41], with brittle materials being below and ductile materials exceeding the critical value. Considering this criterion $Mo_2ScAlC_2$ should behave like a brittle material as its $B/G = 1.65$.

Poisson's ratio $v$ is a very important parameter that provides information regarding the characteristics of the bonding. Poisson's ratio classifies solid materials into two groups: central force solids and non-central force solids [42]. For central force solids, the Poisson's ratio ranges from 0.25 to 0.50. A material is classified as a non-central force solid if its Poisson's ratio is either less than 0.25 or greater than 0.50. The calculated value of $v$ for the new layered compound $Mo_2ScAlC_2$ is 0.25, implying that it should be a central force solid like its isostructural $Mo_2TiAlC_2$ [40]. The Poisson's ratio is also used to quantify the failure mode of solids. The threshold value of this parameter, $v = 0.26$ [43,44] separates the brittle materials from ductile ones. The values less than the threshold value correspond to a brittle material and the values greater than threshold value is associated with a ductile material. In view of Poisson's ratio, $Mo_2ScAlC_2$ is brittle in nature in agreement with the calculated Pugh's ratio. Poisson's ratio is a useful tool to predict the nature of chemical bonding in solids. With $v = 1.0$, a material is classified as a covalent solid, whereas $v = 0.33$ corresponds to a metallic material [45]. $Mo_2ScAlC_2$ is therefore expected to be characterized by a mixture of covalent and metallic properties.

A practically important topic concerns a body that cannot develop the same strain independently of the direction in which stress is applied (elastically anisotropic). In nature, there are no crystalline solids that violate this type of behavior and a deeper understanding of such anisotropic behavior is, important in crystal physics and engineering. An elastic anisotropy factor seeks to quantify how directionally dependent the elastic properties of a system are. Elastic anisotropy also leads to the anisotropy of thermal expansion and microcracks in the crystal [46]. For this reason, it is essential to study the elastic anisotropy to discover the mechanisms that improve the durability of materials. Among different anisotropy factors, the shear anisotropy index owns importance by quantifying the anisotropy in the bonding between atoms in different planes. For hexagonal crystals, this factor is defined by $A = 4C_{44}/(C_{11} + C_{33} - 2C_{13})$ and is coupled with the {100} shear planes between the ⟨011⟩ and ⟨010⟩ directions. To be isotropic, a hexagonal crystal must have anisotropy factor $A = 1$. With $A$-value less than or greater than unity, a crystal exhibits anisotropy in its elastic properties. The amount of deviation from unity quantifies the level of anisotropy possessed by the crystal for its elastic properties. The large $A$-value makes possible the driving force (tangential force) acting on the screw dislocations to progress the cross-slip pinning process [47]. The $A$-value of $Mo_2ScAlC_2$ deviates more than that of $Mo_2TiAlC_2$ from unity, indicating that the elastic anisotropy is profound in Sc-based $Mo_2ScAlC_2$. The cross-slip pining process is significantly easier to promote in $Mo_2ScAlC_2$ but not in $Mo_2TiAlC_2$. Another anisotropy factor defined by the ratio between the linear compressibility coefficients along the $c$- and $a$-axis: $k_c/k_a = (C_{11} + C_{12} - 2C_{13})/(C_{33} - C_{13})$ is calculated. The results indicate that the compressibility along the $c$-axis is slightly smaller than that along the $a$-axis in $Mo_2ScAlC_2$, whereas the compressibility along the c-axis is slightly larger than that along the a-axis for $Mo_2TiAlC_2$. In fact, $Mo_2ScAlC_2$ is more incompressible along the c-axis compared with $Mo_2TiAlC_2$.

*Mechanical wave velocity and Debye temperature*

The elastic moduli of a solid links the Debye temperature and the mechanical wave (sound wave) velocity with which it travels through the solid. The transverse and longitudinal velocity of sound traversing through a crystalline solid with bulk modulus $B$ and shear modulus $G$ can be obtained from [48]:

$$v_t = \left[\frac{G}{\rho}\right]^{1/2} \quad \text{and} \quad v_l = \left[\frac{3B + 4G}{\rho}\right]^{1/2} \tag{3}$$

where $\rho$ refers to the mass-density of the solid. The average sound velocity $v_m$ can be evaluated from the transverse and longitudinal sound velocities using [48]:

$$v_m = \left[\frac{1}{3}\left(\frac{1}{v_l^3} + \frac{2}{v_t^3}\right)\right]^{-1/3} \tag{4}$$

The average sound velocity is subsequently linked to the one of the standard methods to determine the Debye temperature via [48]:

$$\theta_D = \frac{h}{k_B}\left[\left(\frac{3n}{4\pi}\right)\frac{N_A\rho}{M}\right]^{1/3} v_m \tag{5}$$

where $h$ denotes the Planck's constant, $k_B$ is Boltzmann's constant, $N_A$ refers to Avogadro's number, $M$ defines the molecular weight and $n$ is the number of atoms in the molecule.

**Table 3**. Calculated density ($\rho$ in gm/cm$^3$), longitudinal, transverse and average sound velocities ($v_l$, $v_t$, and $v_m$ in km/s) and Debye temperature ($\theta_D$ in K) of Mo$_2$ScAlC$_2$ in comparison with Mo$_2$TiAlC$_2$.

| Compounds | $\rho$ | $v_l$ | $v_t$ | $v_m$ | $\theta_D$ | Ref. |
|---|---|---|---|---|---|---|
| Mo$_2$ScAlC$_2$ | 6.375 | 12.136 | 4.058 | 4.617 | 592.7 | This |
| Mo$_2$TiAlC$_2$ | 6.297 | 8.006 | 4.715 | 3.196 | 413.6 | [31] |

The calculated mechanical wave velocities and Debye temperature are reported in Table 3. The longitudinal and average sound velocities are found to increase, though the transverse sound velocity decreases with the replacement of Ti by Sc. The Debye temperature increases following the average sound velocity as $\theta_D$ is directly proportional to $v_m$. As a general rule, a higher Debye temperature is associated to a higher phonon thermal conductivity. Therefore, Mo$_2$ScAlC$_2$ should be thermally more conductive than Mo$_2$TiAlC$_2$.

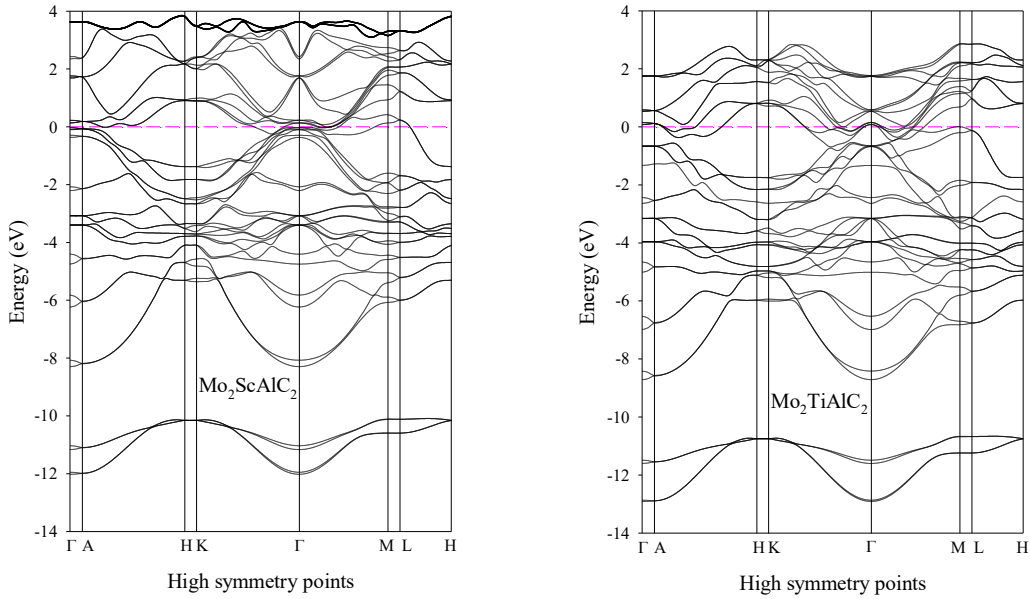

Fig. 2. Electronic band structure of ordered MAX phases Mo$_2$ScAlC$_2$ and Mo$_2$TiAlC$_2$.

### 3.3. Electronic properties

The bonding nature of a solid can be described with its electronic properties i.e., band structure, density of states (DOS), Mulliken atomic populations, total charge density and Fermi surface. The calculated band structure of Mo$_2$ScAlC$_2$ is shown in Fig. 2 along with the band structure of Mo$_2$TiAlC$_2$ [31]. Although the band profiles of these two isostructural ordered MAX phases are similar, there are some distinct differences between them. The metallic bonding exists within both compounds due to the overlapping of conduction bands with valence bands, therefore no band gap is observed at the Fermi level of both nanolaminates. The Fermi surface of Mo$_2$ScAlC$_2$ appears just above the valence band maximum near the Γ point, whereas the Fermi surface of Mo$_2$TiAlC$_2$ becomes visible just below the valence bands maximum close to the Γ point. The conduction bands of Mo$_2$ScAlC$_2$ are wider than those of Mo$_2$TiAlC$_2$. Conversely, the lowest lying valence bands of Mo$_2$ScAlC$_2$ are rather narrow in comparison with those of Mo$_2$TiAlC$_2$. Comparatively, a higher number of valence bands of Mo$_2$ScAlC$_2$ assemble at Fermi level around the Γ point. The overlapping between the valence and conduction bands is more significant in Mo$_2$ScAlC$_2$ compared to Mo$_2$TiAlC$_2$. Therefore, it is expected that Mo$_2$ScAlC$_2$ should be more conductive than Mo$_2$TiAlC$_2$.

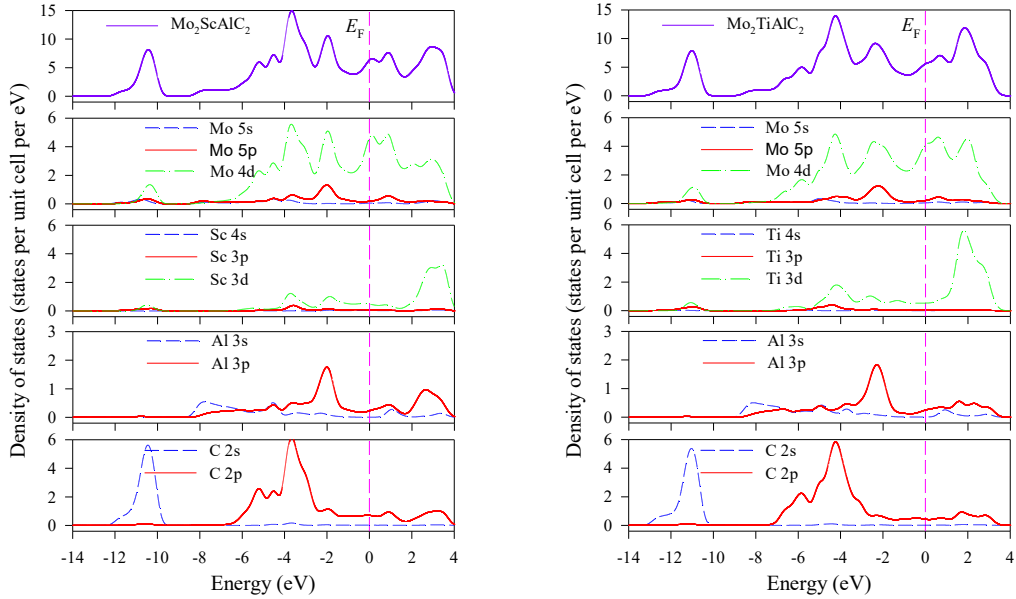

Fig. 3. Density of states of ordered MAX phases Mo$_2$ScAlC$_2$ and Mo$_2$TiAlC$_2$.

The DOS provide insights into the chemical bonding in crystals. The calculated total and partial DOS of Mo$_2$ScAlC$_2$ are shown in Fig. 3 together with those of Mo$_2$TiAlC$_2$ [31] to facilitate comparison. In both phases, the most of the states at the Fermi energy $E_F$ comes from the d-orbitals of the transition metal Mo. These d-resonances at the vicinity of the Fermi level make the two nanolaminates electrically conductive. The calculated total DOS at the Fermi levels $N(E_F)$ are found to be 6.5 and 5.6 eV for Mo$_2$ScAlC$_2$, and Mo$_2$TiAlC$_2$, respectively, indicating that the level of metallic conductivity is expected to be high in Mo$_2$ScAlC$_2$ in comparison to Mo$_2$TiAlC$_2$. The lowest valence band in both ordered MAX phases situated between ~ – 13.1 eV and ~ – 9.5 eV arises due to hybridization between s-orbitals of C and d-orbitals of Mo as well as Sc/Ti. As a result, covalent bonding between transition metals and C is developed. In both phases the Mo–C bond should be stronger than the other bond Sc–C/Ti–C because of high density of Mo d-states. The higher valence band of Mo$_2$ScAlC$_2$ consists of four distinct peaks, whereas the valence band of Mo$_2$TiAlC$_2$ contains three distinct peaks. The additional peak of Mo$_2$ScAlC$_2$ is due to the hybridization of C 2p states with Mo 4d states. The peak at the left of the highest peak in both MAX phases originates from the hybridization between C 2p and Mo 4d states. The highest peak consists of the leading contribution from C 2p and Mo 4d states with a small contribution from Sc/Ti 3d states. The peak at the right of the highest peak arises due to the hybridization of Al 3p orbitals with Mo 4d/5p orbitals. This hybridization leads to the formation of the Mo-Al bond, which is weaker than both Mo–C and Ti-C bonds due to the position near Fermi level. To summarize, it can be concluded that $Mo_2ScAlC_2$ has a mixed character of strong covalent and metallic bonding. Commonly, to most MAX phases it also contains some ionic nature because of the difference in electronegativity between the constituting elements.

*Mulliken atomic population and Vickers hardness*

Mulliken atomic populations are largely based on first-order electron density functions within a linear combination of atomic orbitals-molecular orbital (LCAO- MO) theory [49]. Mulliken population analysis assesses the distribution of electrons in several fractional means among the different parts of atomic bonds. In addition, the overlap population abides a good relation with covalency of bonding and bond strength [50]. Moreover, the overlap population is a convenient way to measure the potency of chemical bonding in DFT calculations [51]. However, Mulliken atomic population analysis is not applicable with the CASTEP code as the DFT method is formulated based on a plane-wave basis set that provides no straightforward way to compute the local atomic properties. Sanchez-Portal *et al*. [52] made CASTEP code to be suitable for Mulliken atomic populations with a technique in which a projection of plane-wave states onto a linear combination of atomic orbitals (LCAO) is used. The atomic charges and the bond populations can be obtained from Mulliken atomic population analysis. The Mulliken charge associated with a particular atom α can be calculated as [49]:

$$Q(\alpha) = \sum_k w_k \sum_\mu^{on\,\alpha} \sum_\nu P_{\mu\nu}(k) S_{\mu\nu}(k) \qquad (6)$$

The overlap population between two atoms α and β can be expressed as [49]:

$$P(\alpha\beta) = \sum_k w_k \sum_\mu^{on\,\alpha} \sum_\nu^{on\,\beta} 2 P_{\mu\nu}(k) S_{\mu\nu}(k) \qquad (7)$$

where $P_{\mu\nu}$ refers to an element of the density matrix and $S_{\mu\nu}$ denotes the overlap matrix. The nature of chemical bonding in crystalline solids can be realized with the knowledge of effective valence charge and Mulliken atomic population. The effective valence charge is estimated from the difference between the formal ionic charge and Mulliken charge on the anion species in a crystal. The identity of a chemical bond either as covalent or ionic with its strength can be determined with the effective valence. The zero effective valences are associated with a purely ionic bond, while the values greater than zero indicate the increasing levels of covalency. The calculated effective valences listed in Table 4 imply that the two isostructural ordered MAX phases include chemical bonding with prominent covalency.

**Table 4.** Population analysis of $Mo_2ScAlC_2$ and $Mo_2TiAlC_2$ [31].

| Compounds | Species | Mulliken Atomic populations | | | | | Effective valence Charge (e) |
|---|---|---|---|---|---|---|---|
| | | S | P | D | Total | Charge (e) | |
| $Mo_2ScAlC_2$ | C | 1.48 | 3.18 | 0.00 | 4.66 | – 0.66 | -- |
| | Al | 0.91 | 1.76 | 0.00 | 2.67 | 0.33 | 2.67 |
| | Sc | 1.99 | 6.36 | 1.69 | 10.04 | 0.96 | 2.04 |
| | Mo | 2.27 | 6.72 | 5.00 | 13.99 | 0.10 | 5.90 |
| $Mo_2TiAlC_2$ | C | 1.45 | 3.19 | 0.00 | 4.64 | – 0.64 | -- |
| | Al | 0.91 | 1.79 | 0.00 | 2.70 | 0.30 | 2.70 |
| | Ti | 2.04 | 6.45 | 2.64 | 11.12 | 0.88 | 2.12 |
| | Mo | 2.23 | 6.70 | 5.02 | 13.95 | 0.05 | 5.95 |

**Table 5.** Calculated Mulliken bond number $n^\mu$, bond length $d^\mu$, bond overlap population $P^\mu$, bond volume $v_b^\mu$ and bond hardness $H_v^\mu$ of $\mu$-type bond and metallic population $P^{\mu'}$ and Vickers hardness $H_v$ of $Mo_2ScAlC_2$ and $Mo_2TiAlC_2$ [31].

| Compounds | Bond | $n^\mu$ | $d^\mu$ (Å) | $P^\mu$ | $P^{\mu'}$ | $v_b^\mu$ (Å$^3$) | $H_v^\mu$ (GPa) | $H_v$ (GPa) |
|---|---|---|---|---|---|---|---|---|
| $Mo_2ScAlC_2$ | Mo–C | 4 | 2.11152 | 1.45 | 0.024 | 8.4525 | 30.09 | 8.03 |
| | Sc–C | 4 | 2.28232 | 0.51 | 0.024 | 10.6741 | 6.95 | |
| | Mo–Al | 4 | 2.78194 | 0.49 | 0.024 | 19.3305 | 2.48 | |
| $Mo_2TiAlC_2$ | Mo-C | 4 | 2.11238 | 1.21 | 0.025 | 8.3452 | 25.54 | 9.01 |
| | Ti-C | 4 | 2.15599 | 0.76 | 0.025 | 8.8727 | 14.30 | |
| | Mo-Al | 4 | 2.79205 | 0.40 | 0.025 | 19.2709 | 2.00 | |

The calculated bond overlap populations for only nearest neighbors in the ordered new MAX phase compound are presented in Table 5. The overlap population of nearly zero value indicates that the interaction between the electronic populations of the two atoms is insignificant and a bond with a smallest Mulliken population is extremely weak and which can be ignored to calculate the materials' hardness. A high degree of ionicity is observed to be associated with a low overlap population. Conversely, a high value is an indication of a high degree of covalency in the chemical bond. The positive and negative bond overlap populations are due to the bonding and antibonding states, respectively. It is observed that both Mo–C and Mo–Al bonds are more covalent in $Mo_2ScAlC_2$ than in $Mo_2TiAlC_2$. The degree of covalency of Sc–C bond is slightly low compared to its similar bond Ti–C.

The prediction of hardness using Mulliken population using DFT is of interest. Gao [53] developed a formula that can successfully calculate the Vickers hardness of non-metallic compounds. Due to having delocalized metallic bonding the metallic and semi-metallic compounds cannot adopt with this formalism [54]. Making a correction due to such bonding Gou et al. [55] proposed an equation for metallic compounds:

$$H_v^\mu = 740(P^\mu - P^{\mu'})(v_b^\mu)^{-5/3} \qquad (8)$$

where $P^\mu$ refers to the Mulliken overlap population of the $\mu$-type bond, $P^{\mu'}$ is the metallic population and is calculated from the cell volume $V$ and the number of free electrons in a cell $n_{free} =$ $\int_{E_P}^{E_F} N(E)dE$ as $P^{\mu'} = n_{free}/V$, and $v_b^\mu$ refers the volume of a bond of $\mu$-type that is evaluated from the bond length $d^\mu$ of type $\mu$ and the number of bonds $N_b^\nu$ of type ν per unit volume as $v_b^\mu = (d^\mu)^3/\sum_\nu[(d^\mu)^3 N_b^\nu]$. The hardness of a complex multiband crystal can be obtained from the geometric average of all individual bond hardness [56,57]:

$$H_V = [\overset{\mu}{\Pi}(H_V^\mu)^{n^\mu}]^{1/\sum n^\mu} \tag{9}$$

where $n^\mu$ is the number of bond of type $\mu$ that composes the real multiband crystals. The calculated Vickers hardness of Mo$_2$ScAlC$_2$ along with its analogue Mo$_2$TiAlC$_2$ is listed in Table 5. The $H_V$ value of 8.03 GPa for Mo$_2$ScAlC$_2$ is calculated to be smaller compared to 9.01 GPa for Mo$_2$TiAlC$_2$. It is clear that the replacement of Ti with Sc reduces the hardness of Mo$_2$ScAlC$_2$ and makes it relatively soft and easily machinable compared to Mo$_2$TiAlC$_2$.

*Fermi surface and charge density*

The Fermi surface topology of Mo$_2$ScAlC$_2$ in the equilibrium structure at *P* = 0 is shown in Fig. 4. Both electron and hole-like sheets are present in the calculated Fermi surface. The central sheet (inner green) is cylindrical and centered along the Γ-A direction of the Brillouin zone. This sheet is surrounded by a very complicated sheet (purple color) having six tubes connected to each other and parallel to the Γ-A direction. Each tube contains an inner tube of reduced radii (pink color). The third sheet is hexagonal, which expands along the Γ-K direction and shrinks along the Γ-M directions. This sheet surrounds the first two sheets. The forth sheet consists of six tuning fork-like parts along the Γ-M directions. The stem (handle) of each tuning fork is situated inside the third sheet. The prongs (U-shaped part) of the tuning fork situated outside the third sheet contain a half-tube cutting along its own axis. These tubes are the parts of the fifth sheet of the Fermi surface. There exist no additional sheets at the corners of the Brillouin zone. The Fermi surface topology is formed mainly due to the low-dispersive Mo 4d-like bands, which should be responsible for the conductivity in Mo$_2$ScAlC$_2$.

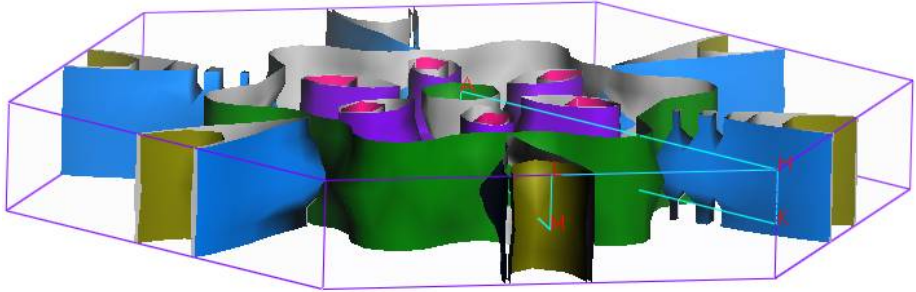

**Fig. 4.** Fermi surface of new quaternary MAX phase carbide Mo$_2$ScAlC$_2$.

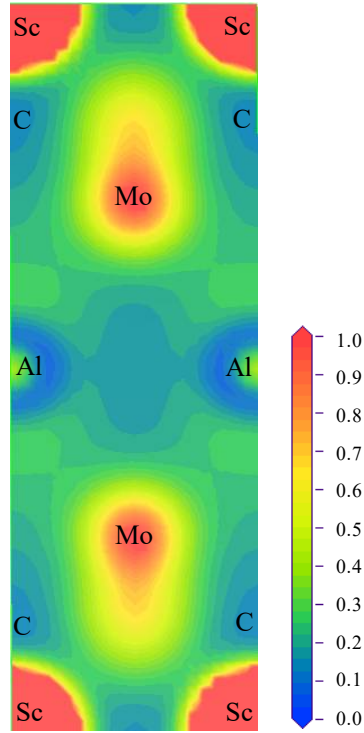

**Fig. 5.** Electronic charge density in (11$\bar{2}$0) plane of Mo$_2$ScAlC$_2$. The scale indicates the low and high electron density with blue and red color, respectively.

The electron charge density distribution of Mo$_2$ScAlC$_2$ is investigated to obtain further insights into the chemical bonding. The contour of calculated electron charge density in the (11$\bar{2}$0) plane is shown in Fig. 5. The atom of high electronic charge i.e., electronegativity pulls electron density towards itself [58]. The electron charge density map discloses a strong directional Mo–C–Mo covalent bond chain with each pair of the chain coupled with a comparatively weak Sc–C bond. Due to a large variation in electronegativity, the electronic charge in the region of Mo atoms is attracted towards C atoms. Consequently, a strong covalent-ionic bonding along Mo and C direction is induced. This bond arises due to hybridization between C 2s and Mo 4d electrons. Relatively weak Mo-Al bonds form as the electron charge density of Mo atom just overlaps with that of the Al atom. The present results are consistent with the findings that MAX phases typically have reasonably strong M–X bonds and relatively weak M–A bonds [3].

*3.4. Intrinsic defect processes*

*Frenkel defect formation*

The energetics of intrinsic defect process and in particular the Frenkel defects can be important to access the radiation tolerance of materials. Considering for example nuclear applications a low pair formation energy may be linked with a higher content of more persistent defects. These in turn may lead to the loss of ordering in the crystal. The following relations are the key Frenkel reactions in Kröger–Vink notation (here $V_A$ and $A_i$ denote a vacant A site and an A interstitial defect respectively) [59]:

$$M_M \rightarrow V_M + M_i \tag{10}$$

$$A_A \rightarrow V_A + A_i \tag{11}$$

$$X_X \rightarrow V_X + X_i \tag{12}$$

As it has been previously [60] discussed in 312 MAX phases there are numerous possible interstitial sites. The calculated energetically preferable interstitials sites are: $Mo_i$ (3/4, 0.636, 1/4), $Sc_i$ (0.764, 0.699, 1/4), $Al_i$ (2/3, 1/3, 1/4) and $C_i$ (0.340, 2/3, 1/4).

*Antisite defect formation*

During radiation damage the produced point defects may either recombine or reside on alternative lattice site forming antisite defects [61]. From a physical viewpoint a low energy antisite formation energy means that a high proportion of residual defects will remain in the material as typically the conversion of an interstitial into an antisite will result to a net reduction of defect mobility [61]. The antisite formation mechanisms are:

$$M_M + A_A \rightarrow M_A + A_M \tag{13}$$

$$M_M + X_X \rightarrow M_X + X_M \tag{14}$$

$$A_A + X_X \rightarrow A_X + X_A \tag{15}$$

**Table 6.** The calculated defect reaction energies (in eV, for relations 10-15) for the $Mo_2ScAlC_2$ MAX phase.

| Reaction | Frenkel | Reaction | Antisite |
|---|---|---|---|
| $Mo_{Mo} \rightarrow V_{Mo} + Mo_i$ | 6.393 | $Mo_{Mo} + Al_{Al} \rightarrow Mo_{Al} + Al_{Mo}$ | 2.866 |
| $Sc_{Sc} \rightarrow V_{Sc} + Sc_i$ | 8.494 | $Mo_{Mo} + C_C \rightarrow Mo_C + C_{Mo}$ | 9.106 |
| $Al_{Al} \rightarrow V_{Al} + Al_i$ | 4.656 | $Sc_{Sc} + Al_{Al} \rightarrow Sc_{Al} + Al_{Sc}$ | 2.835 |
| $C_C \rightarrow V_C + C_i$ | 2.503 | $Sc_{Sc} + C_C \rightarrow Sc_C + C_{Sc}$ | 11.995 |
|  |  | $Al_{Al} + C_C \rightarrow Al_C + C_{Al}$ | 8.386 |

*Implications of defect processes*

The radiation performance of materials is dependent upon their propensity to form and accommodate point defects [60,61]. A high concentration of defects may lead to the destabilization of the MAX phase (or any material) and may result to volume changes and microcracking [61-63]. It is established that displacive radiation can result to an athermal concentration of Frenkel pairs and therefore the radiation tolerance of materials is linked to their ability to resist the formation of significant populations of Frenkel and/or antisite) defects. Therefore, high Frenkel and antisite defect energies may be considered as a condition of radiation tolerance.

As it can be observed from Table 6 which is based on the defect processes investigated by DFT the dominant intrinsic defect mechanism is the carbon Frenkel energy (2.503 eV). Additionally, antisite processes could lead to the formation of $Mo_{Al} + Al_{Mo}$ (2.866 eV) and $Sc_{Al} + Al_{Sc}$ (2.835 eV). It can be concluded from these results that other MAX phases such as $Ti_3AlC_2$ [60,61] is more radiation tolerant than $Mo_2ScAlC_2$.

## 4. Concluding remarks

To summarize, the mechanical behavior, bonding nature and defect processes of $Mo_2ScAlC_2$ is calculated using DFT for the first time. To facilitate comparison we consider the isostructural $Mo_2TiAlC_2$ MAX phase. The relaxed structural parameters are in good agreement with the experimental results. The calculated single crystal elastic constants ensure the mechanical stability of the compound. $Mo_2ScAlC_2$ is calculated to be prone to shear along the crystallographic *b* and *c* axes, when an eligible force is applied to the crystallographic *a* axis. Under uniaxial stress, the compressibility along the ⟨001⟩ direction is predicted to be easier in $Mo_2ScAlC_2$. The volume deformation is expected to be easier in $Mo_2ScAlC_2$ than in $Mo_2TiAlC_2$. The new compound should behave in a brittle manner. The high Debye temperature of $Mo_2ScAlC_2$ indicates its higher thermal conductivity than $Mo_2TiAlC_2$. The cross-slip pining process is easier to promote in $Mo_2ScAlC_2$ than in $Mo_2TiAlC_2$. The chemical bonding in $Mo_2ScAlC_2$ is a mixture of strong covalent and metallic with little ionic nature. Bonds Mo–C and Mo–Al should be more covalent in $Mo_2ScAlC_2$ than those in $Mo_2TiAlC_2$. The degree of covalency of Sc–C bond is rather low compared to the Ti–C bond. Due to its low hardness $Mo_2ScAlC_2$ is expected to be softer and more easily machinable compared to $Mo_2TiAlC_2$. The low-dispersive Mo 4d-like bands should be responsible to form the Fermi surface topology of the new compound. The carbon Frenkel defect reaction is the lowest energy intrinsic defect process in $Mo_2ScAlC_2$. The level of radiation tolerance in $Mo_2ScAlC_2$ is evaluated to be relatively low compared to other MAX phases such as $Ti_3AlC_2$.

**Acknowledgements**
S.R.G.C. and A.C. are grateful for funding from the Lloyd's Register Foundation, a charitable foundation helping to protect life and property by supporting engineering-related education, public engagement and the application of research.

**References**
[1] W. Jeitschko, H. Nowotny and F. Benesovsky, Monatsh. Chem. 94 (1963) 672.
[2] M.W. Barsoum and T. El-Raghy, J. Am. Ceram. Soc. 79 (1996) 1953.
[3] M. W. Barsoum, Prog. Solid State Chem. 28 (2001) 201.

[4]  H. Yoo, M. W. Barsoum and T. El-Raghy, Nature (London) 407 (2000) 581.

[5]  T. El-Raghy *et al.*, J. Am. Ceram. Soc. 82 (1999) 2855.

[6]  M.W. Barsoum, L. Farber and T. El-Raghy, Metall. Mater. Trans. A 30 (1999) 1727.

[7]  Z. M. Sun *et al.*, Mater. Trans. 47 (2006) 170.

[8]  P. Finkel, M. W. Barsoum and T. El-Raghy, J. Appl. Phys. 87 (2000) 1701.

[9]  M. W. Barsoum, Physical properties of the MAX phases, *Encyclopedia of Materials: Science and Technology* (Elsevier, Amsterdam, 2009).

[10] M. Radovic *et al.*, J. Alloys Compds. 361 (2003) 299.

[11] C. J. Gilbert *et al.*, Scr. Mater. 238 (2000) 761.

[12] M. Sundberg *et al.*, Ceram. Int. 30 (2004) 1899.

[13] D. Horlait, S. Grasso, A. Chroneos, and W. E. Lee, Mater. Res. Lett. 4 (2016) 317.

[14] S. E. Lofland, J. D. Hettinger, K. Harrell, P. Finkel, S. Gupta, M. W. Barsoum, and G. Hug, Appl. Phys. Lett. 84 (2004) 508.

[15] D. Horlait, S. C. Middleburgh, A. Chroneos and W. E. Lee, Sci. Rep. 6 (2016) 18829.

[16] R. Meshkian, Q. Tao, M. Dahlqvist, J. Lu, L. Hultman, and J. Rosen, Acta Materialia 125 (2017) 476.

[17] H. Nowotny, Strukturchemie einiger verbindungen der übergangsmetalle mitden elementen C, Si, Ge, Sn, Prog. Solid. State. Chem. 5 (1971) 27.

[18] L.E. Toth, W. Jeitschko, C.M. Yen, The superconducting behavior of several complex carbides and nitrides, J. Less Common Metals 10 (1966) 29.

[19] S. J. Clark, M. D. Segall, C. J. Pickard, P. J. Hasnip, M. I. J. Probert, K. Refson, and M. C. Payne, Zeitschrift für Kristallographie 220 (2005) 567.

[20] J. P. Perdew, K. Burke, and M. Ernzerhof, Phys. Rev. Lett. **77** (1996) 3865.

[21] D. Vanderbilt, Phys. Rev. B 41(1990) 7892.

[22] H. J. Monkhorst, J. D. Pack, Phys. Rev. B 13 (1976) 5188.

[23] T. H. Fischer and J. Almlof, J. Phys. Chem. 96 (1992) 9768.

[24] F. D. Murnaghan, Finite Deformation of an Elastic Solid (Wiley, New York, 1951).

[25] M. A. Hadi, M. Roknuzzaman, F. Pervin, S. H. Naqib, A. K. M. A. Islam and M. Aftabuzzaman, J. Sci. Res. 6 (2014) 11.

[26] M. T. Nasir, M. A. Hadi, S. H. Naqib, F. Pervin, A. K. M. A. Islam, M. Roknuzzaman, and M. S. Ali, Int. J. Mod. Phys. B 28 (2014) 1550022.

[27] M. Roknuzzaman, M. A. Hadi, M. J. Abden, M. T. Nasir, A. K. M. A. Islam, M. S. Ali, K. Ostrikov, and S. H. Naqib, Comp. Mater. Sci. 113 (2016) 146.

[28] M. A. Hadi, Comp. Mater. Sci. 117 (2016) 422.

[29] M. A. Hadi, M. A. Alam, M. Roknuzzaman, M. T. Nasir, A. K. M. A. Islama, and S. H. Naqib, Chin. Phys. B 24 (2015) 117401.

[30] M. A. Alam, M. A. Hadi, M. T. Nasir, M. Roknuzzaman, F. Parvin, M. A. K. Zilani, A. K. M. A. Islam, S. H. Naqib, J. Supercond. Nov. Magn. 29 (2016) 2503.

[31] M. A. Hadi and M. S. Ali, Chin. Phys. B 25 (2016) 107103.

[32] M. A. Hadi, M. T. Nasir, M. Roknuzzaman, M. A. Rayhan, S. H. Naqib, and A. K. M. A. Islam, Phys. Status Solidi B, 253 (2016) 2020.

[33] M. A. Hadi, M. S. Ali, S. H. Naqib and A. K. M. A. Islam, Chinese Physics B 26 (2017) (acceptd).

[34] M. H. K. Rubel, M. A. Hadi, M. M. Rahaman, M. S. Ali, M. Aftabuzzaman, R. Pervin, A. K. M. A. Islam, and N. Kumada, Comp. Mater. Sci. (2017) (submitted).

[35] W. Voigt, *Lehrbuch der Kristallphysik*, (Taubner, Leipzig, 1928).

[36] A. Reuss and Z. Angew, Math. Mech. 9 (1929) 49.

[37] R. Hill, Proc. Phys. Soc. A 65 (1952) 349.

[38] B. Anasori, J. Halim, J. Lu, A. Cooper, Voigt, L. Hultman and M. W. Barsoum, Scripta Materialia 101 (2015) 5.

[39] M. Born, Math. Proc. Cambridge Philos. Soc. 36 (1940) 160.

[40] B. Anasori, M. Dahlqvist, J. Halim, E. J. Moon, J. Lu, B. C. Hosler, E. N. Caspi, S. J. May, L. Hultman, P. Eklund, J. Rose′n, and M. W. Barsoum, J. Appl. Phys. 118 (2015) 094304.

[41] S.F. Pugh, Phil. Mag. 45 (1954) 823.

[42] O. L. Anderson and H. H. Demarest Jr., J. Geophys. Res. 76 (1971) 1349.

[43] I. N. Frantsevich, F. F. Voronov, and S.A. Bokuta, Elastic constants and elastic moduli of metals and insulators handbook, (Naukova Dumka, Kiev, 1983, pp. 60–180).

[44] G. Vaitheeswaran, V. Kanchana, A. Svane and A. Delin, J. Phys.: Conden. Matter 19 (2007) 326214.

[45] A. Savin, D.C.H. Flad, J. Flad, H. Preuss, H.G. Schnering, Angew. Chem. Int. Ed. Engl. 31 (1992) 185.

[46] V. Tvergaard and J. W. Hutchinson, J. Am. Chem. Soc. 71 (1988) 157.

[47] M. H. Yoo, Scr. Metall. 20 (1986) 915.

[48] E. Schreiber, O. L. Anderson, N. Soga, Elastic Constants and Their Measurements (McGraw-Hill, New York, 1973)

[49] R. S. Mulliken, J. Chem. Phys. 23 (1955) 1833.

[50] M. D. Segall *et al.*, Phys. Rev. B 54 (1996) 16317.

[51] W. Y. Ching and Y. N. Xu, Phys. Rev. B 59 (1999) 12815.

[52] D. Sanchez-Portal, E. Artacho, and J. M. Soler, Solid State Commun. 95 (1995) 685.

[53] F. M. Gao, Phys. Rev. B 73 (2006) 132104.

[54] J. H. Westbrook and H. Conrad, The Science of Hardness Testing and Its Research Applications (ASM, Ohio, 1973).

[55] H. Gou, L. Hou, J. Zhang and F. Gao, Appl. Phys. Lett. 92 (2008) 241901.

[56]   A. Szymański and J. M. Szymański, Hardness Estimation of Minerals Rocks and Ceramic Materials, 6$^{th}$ edition, (Elsevier, Amsterdam, 1989).

[57]   V. M. Glazov and V. N. Vigdorovid, Hardness of Metals (Izd. Metellurgiya, Moskva, 1989).

[58]   IUPAC 1997 Compendium of Chemical Technology, 2nd edn. (the "Gold Book"); 2006 Online corrected version.

[59]   F. A. Kröger and H. J. Vink, Solid State Phys. 3 (1956) 307.

[60]   E. Zapata-Solvas, S.-R. G. Christopoulos, N. Ni, D. C. Parfitt, D. Horlait, M. E. Fitzpatrick, A. Chroneos, and W. E. Lee, J. Am. Ceram. Soc. DOI: 10.1111/jace.14742

[61]   S. C. Middleburgh, G. R. Lumpkin, and D. Riley, J. Am. Ceram. Soc. 96 (2013) 3196.

[62]   W. J. Weber, Radiat. Eff. 77 (1983) 295.

[63]   F. W. Clinard Jr, D. L. Rohr, and R. B. Roof, Nucl. Instrum. Methods Phys. Res. B 1 (1984) 581.